\documentclass[11pt]{article}
\usepackage{graphicx,epsfig}

\newcommand{\BABARPubYear}    {04}

\newcommand{\BABARProcNumber} {020}
\newcommand{\SLACPubNumber} {10455}
\newcommand{\LANLNumber} {0000}
\def\sss{\scriptscriptstyle}
\def\barpd{{\raise.35ex\hbox{${\sss (}$}}--{\raise.35ex\hbox{${\sss )}$}}}
\def\dbarp{\hbox{$D^{0}$\kern-1.3em\raise1.5ex\hbox{\barpd}}}

\input pubboard/babarsym
\def\BABAR{\babar}

\long\def\inst#1{\par\nobreak\kern 4pt\nobreak
    {\it #1}\par\vskip 10pt plus 3pt minus 3pt}

\begin{document}
{\pagestyle{empty}

\begin{flushright}
SLAC-PUB-\SLACPubNumber \\
\babar-PROC-\BABARPubYear/\BABARProcNumber \\
hep-ex/\LANLNumber \\
May, 2004 \\
\end{flushright}

\par\vskip 4cm

\begin{center}
\Large \bf B decays to charm states at \Lbabar.
\end{center}
\bigskip

\begin{center}
\large 
G. Calderini\\
Dipartimento di Fisica dell'Universit\`a degli Studi di Pisa e\\
Istituto Nazionale di Fisica Nucleare, Sez. di Pisa\\
via Buonarroti 2, 56100 Pisa, Italy.\\
~\\
(for the \lbabar\ Collaboration)
\end{center}
\bigskip \bigskip

\begin{center}
\large \bf Abstract
\end{center}
In this paper recent results in the field of $B$ meson decays to states 
containing charm are presented. These analyses are based on the 1999-2003 dataset 
collected by the \lbabar experiment
at the PEP-II $e^+e^-$ storage ring at the Stanford Linear Accelerator Center.  
Special attention is devoted to $B$ decays to final states containing $D$ mesons. 
In addition a few new results from $B$ decays to charmonium states are reported. 

\vfill
\begin{center}
Contributed to the Proceedings of the XXXIX$^{th}$ Rencontres de Moriond, QCD and High Energy Hadronic Interactions \\
3/28/2004---4/4/2004, La Thuile, Italy
\end{center}

\vspace{1.0cm}
\begin{center}
{\em Stanford Linear Accelerator Center, Stanford University, 
Stanford, CA 94309} \\ \vspace{0.1cm}\hrule\vspace{0.1cm}
Work supported in part by Department of Energy contract DE-AC03-76SF00515.
\end{center}

\section{Introduction}

The decay of $B$ mesons to states containing charm and charmonium provides 
an excellent laboratory for the study of hadronic $B$ decays.
With more than 200 millions of $B$ pairs presently collected, the \BABAR experiment 
has integrated an unprecedented luminosity, allowing the test of $B$ decay models 
in more channels and with greater precision than ever before, and the observation
of $B$ decays modes never seen in the past. In this paper we present some examples. 
In Section 2 new results from analyses of $B$ decays to modes containing $D$
mesons are presented. Section 3 is devoted to $B$ decays to charmonium states.     
The \BABAR detector is described in detail elsewhere \cite{BaBar_NIM}.
Charge conjugation is implied throughout this note.

\section{$B$ decays to states containing $D$ mesons}
\subsection{Study of the decay $B \rightarrow D_{sJ}D^{(*)}$}

In 2003, the unexpected observation by \BABAR \cite{palano} of a narrow $D_s^+ \pi^0$ 
resonance with a mass of $2317$ MeV/$c^2$ created some excitement. The discovery 
was soon confirmed by CLEO and Belle. In addition CLEO found a second new state 
near $2460$ MeV/$c^2$ in $D_s^{*+} \pi^0$ combinations~\cite{palano_cleo}, soon
confirmed by \BABAR and Belle, who added the decays into $D_s^+ \gamma$ and
$D_s^+ \pi^+\pi^-$. 
A recent analysis by \BABAR was aimed to study the production of $D_{sJ}$ in $B$ decays,
in order to use the kinematical constraint to extract additional information
on the $D_{sJ}$ spin from the angular distribution of the decay products.  
The analysis is based on the data collected during \BABAR Run 1, 2 
and 3, which is equivalent to about 125 millions of $B$ pairs. 
Production of $D_{sJ}(2317)$ and $D_{sJ}(2460)$ has been analyzed in both 
charged and neutral $B$ decays, through the decays $B^+ \rightarrow D_{sJ}^+ D^{(*)0}$
and $B^0 \rightarrow D_{sJ}^+ D^{(*)-}$.   
The $D_{sJ}(2317)$ candidates are reconstructed 
using $D_s^+ \pi^0$ combinations, while $D_{sJ}(2460)$ are reconstructed in the 
$D_s^{*+}\pi^0$ and $D_s^{+}\gamma$ modes. The $D_{sJ}$ candidate are then paired
with a $D$ or $D^*$ meson. 
The analysis has shown a signal in all the three $D_{sJ}$ modes (see Fig. \ref{dsj_plot})  
and preliminary branching fractions for the twelve decay channels are extracted
(see \mbox{Tab 1}).
\begin{figure}[htb]
\begin{center}
\mbox{\epsfig{file=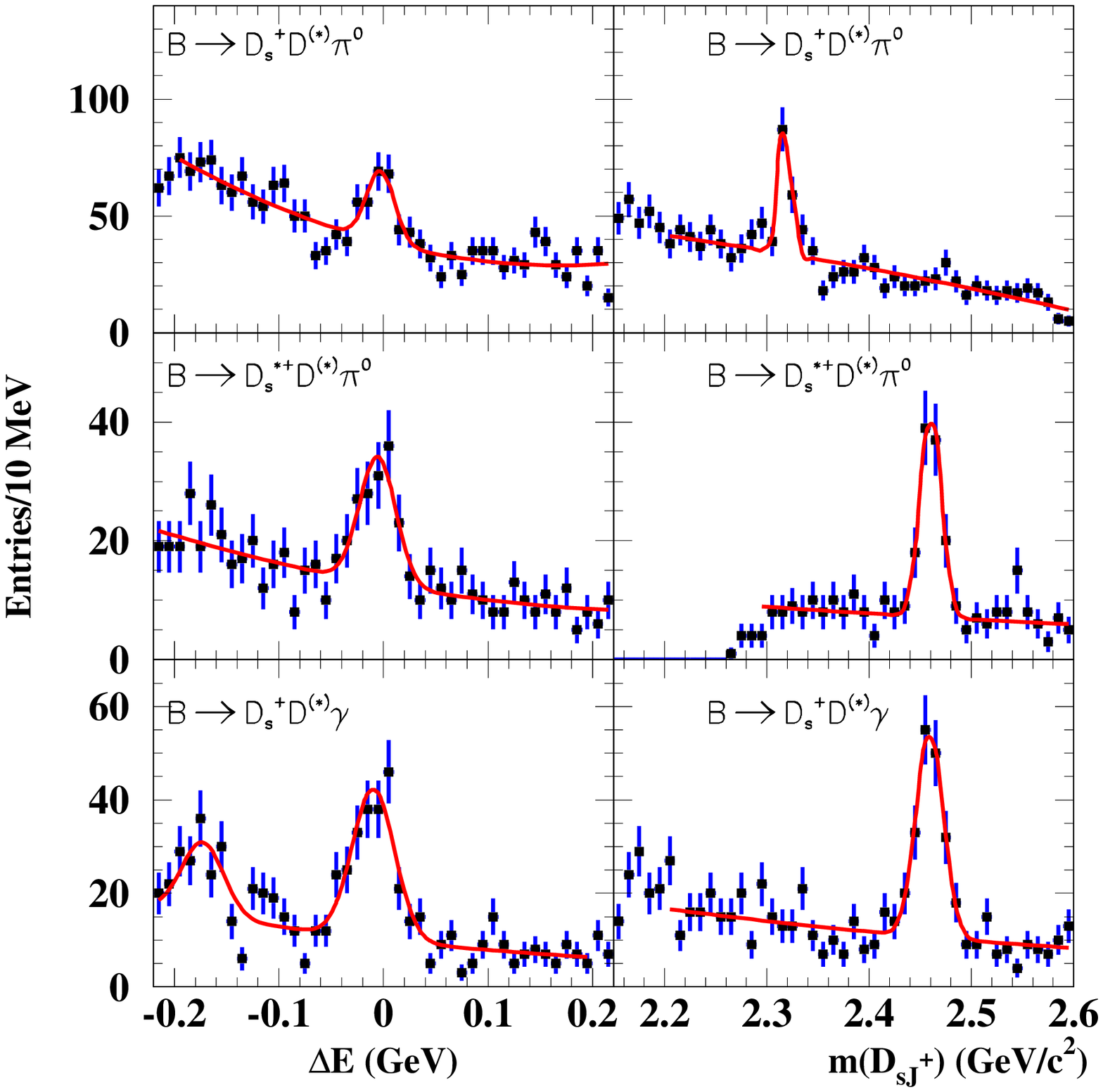,height=8.5cm}}
\end{center}
\caption{$\Delta E$ and $m(D_{sJ})$ spectra for (a,b) the sum of 
$B \rightarrow D_{s0}^+ \overline{D}^{(*)}, D_{s0}^+ \rightarrow D_s \pi^0$;
(c,d) the sum of \mbox{$B \rightarrow D_{s1}^+ \overline{D}^{(*)}, D_{s1}^+ \rightarrow D_s \pi^0$};
(e,f) the sum of \mbox{$B \rightarrow D_{s1}^+ \overline{D}^{(*)}, D_{s1}^+ \rightarrow D_s \gamma$}}
\label{dsj_plot}
\end{figure}
\begin{table}[h!]
\centering
\vspace{2mm}
\begin{tabular}{|c|l|c|c|}
  \hline
 & decay mode \hfill\    &\multicolumn{2}{|c|}{BR/$(10^{-3})$}\\
 && this analysis & Belle \\\hline
  \hline
I&  $B^0\rightarrow D_{s0}^{+}D^{-}$ $(D_{s0}^{+}\rightarrow  D_{s}^{+} \pi^0)$  &
 $2.09 \pm 0.40\pm 0.34 ^{+0.70}_{-0.42}$  & $0.86\pm0.26{}^{+0.33}_{-0.26}$\\
II &  $B^0\rightarrow D_{s0}^{+}D^{*-}$ $(D_{s0}^{+}\rightarrow
D_{s}^{+} \pi^0)$
& $1.12 \pm 0.38\pm 0.20^{+0.37}_{-0.22}  $ & \ \hfill --- \hfill\ \\
III &  $B^+\rightarrow D_{s0}^{+}\overline{D}^{0}$
$(D_{s0}^{+}\rightarrow  D_{s}^{+} \pi^0)$
& $1.28 \pm 0.37\pm 0.22 ^{+0.42}_{-0.26}$ & $0.81\pm0.24{}^{+0.30}_{-0.27}$ \\
IV &  $B^+\rightarrow D_{s0}^{+}\overline{D}^{*0}$
$(D_{s0}^{+}\rightarrow  D_{s}^{+} \pi^0)$
& $1.91 \pm 0.84\pm 0.50 ^{+0.63}_{-0.38}$ & \ \hfill --- \hfill\ \\
  \hline
V & $B^0\rightarrow D_{s1}^{+}D^{-}$ $(D_{s1}^{+}\rightarrow  D_{s}^{*+} \pi^0)$
& $1.71 \pm 0.72\pm 0.27 ^{+0.57}_{-0.35}$ & $2.27\pm0.68{}^{+0.73}_{-0.62}$\\
VI & $B^0\rightarrow D_{s1}^{+}D^{*-}$ $(D_{s1}^{+}\rightarrow
D_{s}^{*+} \pi^0)$
& $5.89 \pm 1.24 \pm 1.16^{+1.96}_{-1.17} $ & \ \hfill --- \hfill\ \\
VII&  $B^+\rightarrow D_{s1}^{+}\overline{D}^{0}$
$(D_{s1}^{+}\rightarrow  D_{s}^{*+} \pi^0)$
& $2.07 \pm 0.71\pm 0.45 ^{+0.69}_{-0.41}$ & $1.19\pm0.36{}^{+0.61}_{-0.49}$\\
VIII&  $B^+\rightarrow D_{s1}^{+}\overline{D}^{*0}$
$(D_{s1}^{+}\rightarrow  D_{s}^{*+} \pi^0)$
& $7.30 \pm 1.68 \pm 1.68^{+2.40}_{-1.43}$ & \ \hfill --- \hfill\ \\
  \hline
IX&  $B^0\rightarrow D_{s1}^{+}D^{-}$ $(D_{s1}^{+}\rightarrow
D_{s}^{+} \gamma)$
& $0.92 \pm 0.24\pm 0.11 ^{+0.30}_{-0.19}$ & $0.82\pm0.25{}^{+0.22}_{-0.19}$\\
X&  $B^0\rightarrow D_{s1}^{+}D^{*-}$ $(D_{s1}^{+}\rightarrow
D_{s}^{+} \gamma)$
& $2.60 \pm 0.39   \pm 0.34^{+0.86}_{-0.52}$ & \ \hfill --- \hfill\ \\
XI&  $B^+\rightarrow D_{s1}^{+}\overline{D}^{0}$
$(D_{s1}^{+}\rightarrow  D_{s}^{+} \gamma)$
& $0.80 \pm 0.21\pm 0.12 ^{+0.26}_{-0.16}$ & $0.56\pm0.17{}^{+0.16}_{-0.15}$\\
XII&  $B^+\rightarrow D_{s1}^{+}\overline{D}^{*0}$
$(D_{s1}^{+}\rightarrow  D_{s}^{+} \gamma)$
& $2.26 \pm 0.47\pm 0.43 ^{+0.74}_{-0.44}$ & \ \hfill --- \hfill\ \\
  \hline
\end{tabular}
\caption{Branching fractions measured in the \BABAR analysis and corresponding Belle results}
\label{tab:dsj_br}
\end{table}
\vspace{0.5cm}

\noindent
The decay modes involving combinations of the $D_{sJ}$ and a neutral or charged $D^*$ are 
first observations. The modes with $D_{sJ}$ and a neutral or charged $D$
were also observed by Belle \cite{Krokovny:2003zq} and the results of the two experiments
are in rather good agreement.
  
The spin of the $D_{sJ}(2460)$ has been investigated using the $D_s \gamma$ decay.
The distribution of the angle $\Theta_h$ between the $D_{sJ}$ flight direction and the $D_s$ 
momentum in the $D_{sJ}$ reference frame has been studied. The distribution 
favours the $J=1$ spin hypothesis over the $J=2$ (Fig. \ref{fig:hel_helpl}).
Spin J=0 is ruled out by parity and angular momentum conservation.   
A similar conclusion has been reached by Belle \cite{Krokovny:2003zq}. 

\begin{figure}[h]
\vspace*{5mm}
\begin{center}
  \hspace*{-0mm}\includegraphics[width=5.5cm]{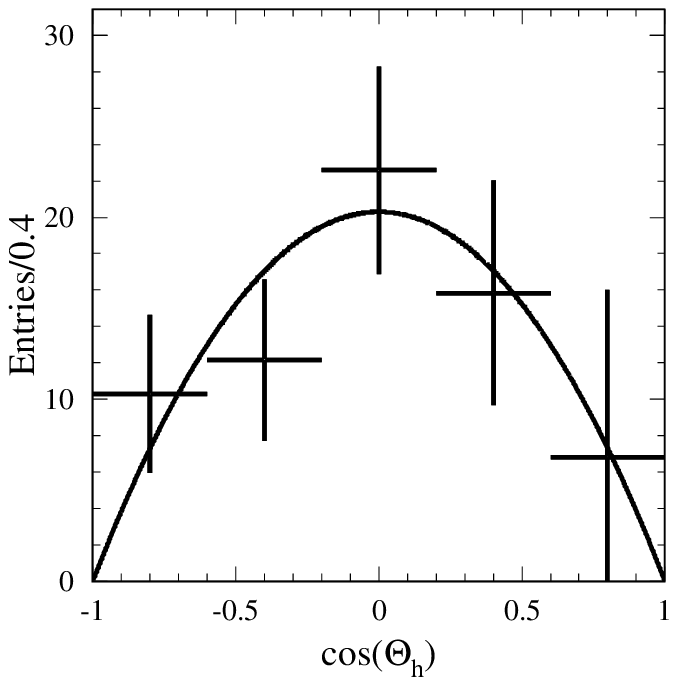}
  \hspace*{-0mm}\includegraphics[width=5.5cm]{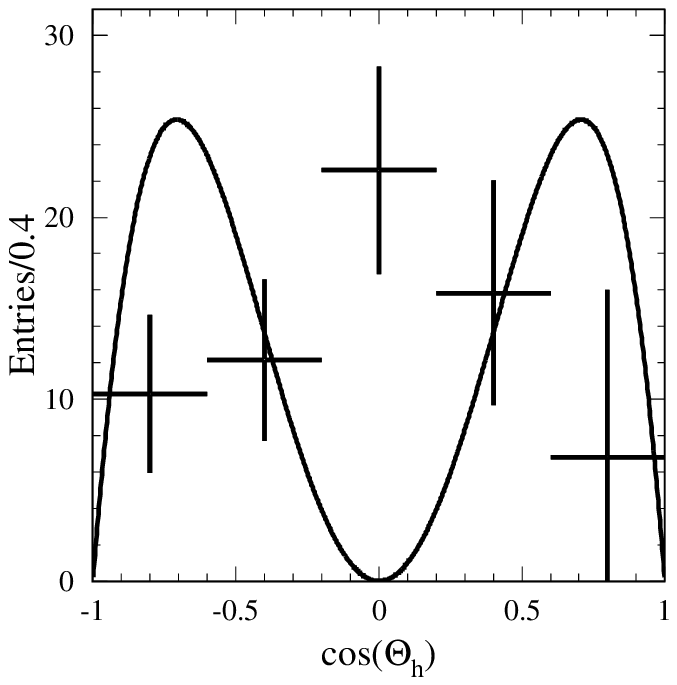}
\end{center}
\caption{Helicity distributions obtained from $m(D_s\gamma)$ fits in the corres
ponding  $\cos(\Theta_h)$ region for selected BaBar data. 
The solid curves are the analytical expectations for two different $D_{sJ}^*(2460)^+$ 
spin hypotheses, which have been normalized to the data: (left) $J=1$ and (right) $J=2$.}
\label{fig:hel_helpl}
\end{figure}
\subsection{Study of the decay $B \rightarrow DK$}
The $B^+$ decay to $\dbarp K^+$ state is a crucial mode for the extraction of the $\gamma$ angle 
of the Unitarity Triangle. 
The original method, suggested by Gronau and Wyler \cite{gronau}, was based on the
interference between the $B \rightarrow D^0 K$ ($b \rightarrow c$) and the 
$B \rightarrow \overline{D}^0K$ ($b \rightarrow u$) diagrams once the $D^0$ and  
the $\overline{D}^0$ mesons decay to CP states 
($K^+K^-$,$\pi^+\pi^-$,$K_s\pi^0$ etc).  
The two diagrams for the two $B$ decays are shown in Fig. \ref{fig:feyn-diag}.
The limitation of this method is that the branching fractions for $D$ decays to CP modes are 
rather small and the interference is even smaller, since the contribution of the 
($b \rightarrow u$) diagram is suppressed with 
respect to the ($b \rightarrow c$) one. 
More recently a variation of the method has been proposed by Atwood, Dunietz and Soni 
(ADS)\cite{atwood}. It is based on the separate measurement of $B^+$ and $B^-$ to a final
state which can be reached by primarily two amplitudes, each of the same order of magnitude.
One amplitude is from a doubly suppressed decay $B^+ \rightarrow \overline{D}^0 K^+$
combined with a favored $\overline D$ decay $\overline D^0 \rightarrow K^+\pi^-$;
the other amplitude is from a favored decay $B^+ \rightarrow D^0 K^+$, followed by a 
doubly suppressed $D$ decay $D^0 \rightarrow K^+\pi^-$. In this way the interference 
is enhanced. 

\begin{figure}[h]
\begin{center}
\epsfig{file=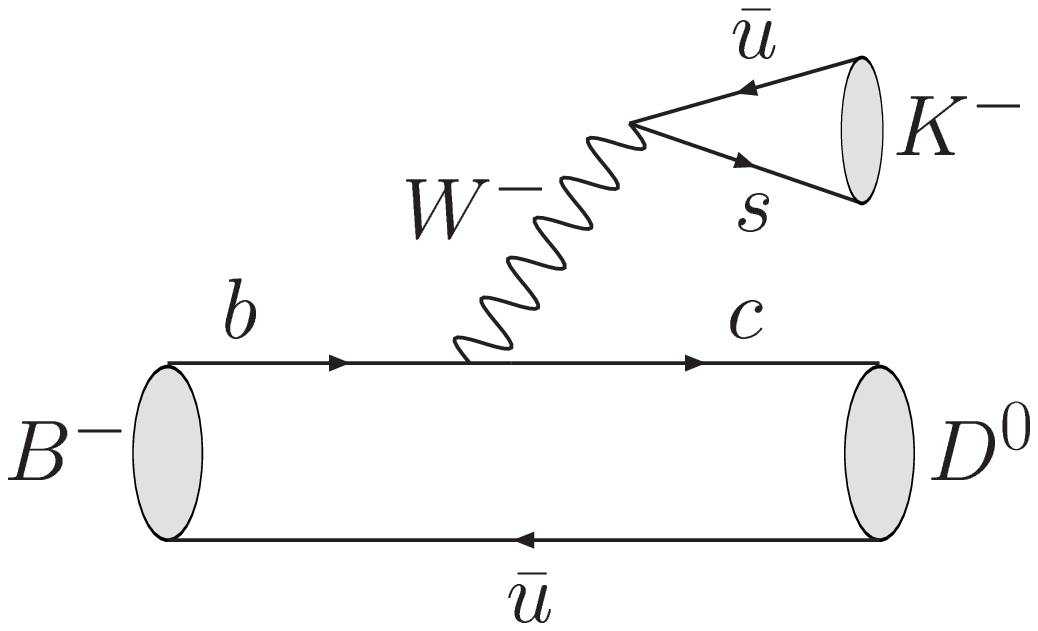,height=3.5cm}
\epsfig{file=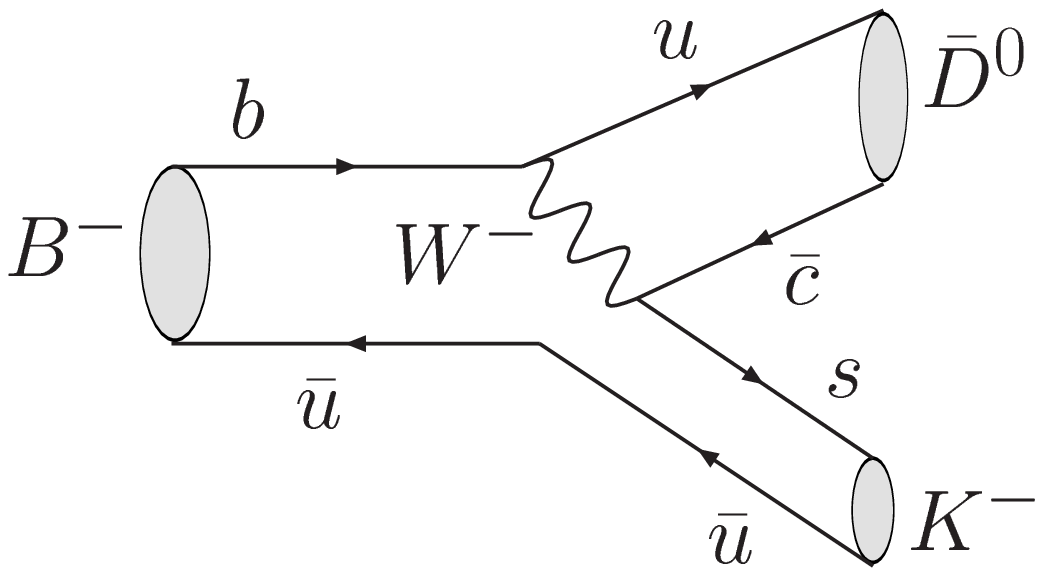,height=3.5cm}
\end{center}
\caption{ The two contributing $B$-decay Feynman diagrams for 
  $B^- \rightarrow D K^-$. 
  The diagram on the right, which has a $b\rightarrow u$ transition,
  is color suppressed, while the $b\rightarrow c$ diagram on the left
  is not.
  }
\label{fig:feyn-diag}
\end{figure}

\noindent Following the ADS method \cite{atwood} it is possible to write:
\begin{equation}
R_{DCS} \equiv 
{{\Gamma(B^- \rightarrow D_{K^+\pi^-})+\Gamma(B^+ \rightarrow D_{K^-\pi^+})} 
\over
{\Gamma(B^- \rightarrow D_{K^-\pi^+})+\Gamma(B^+ \rightarrow D_{K^+\pi^-})} }=
r_D^2+r_B^2+2r_Dr_B\cos{\gamma}\cos{\delta}
\label{eq:r_dcs}
\end{equation}
where $r_D=|A(D^0\rightarrow K^+\pi^-)|/|A(\overline{D}^0\rightarrow K^+\pi^-)|\approx 0.060 \pm 0.003$, 
$r_B=|A(b\rightarrow u)|/|A(b \rightarrow c)|$, and $\delta$ is an overall strong phase difference.

Based on this motivation, an analysis has been performed at \BABAR to study the
decay of $B^+$ to $\dbarp K^+$. The $D$ candidates are reconstructed 
in $K^{\pm} \pi^{\mp}$ modes. The study is based on data collected in Run 1,2,3. 
In spite of the fact that the method is both theoretically and experimentally
very clean, the strong overall suppression of the two amplitudes and the required
level of background rejection still represents a challenge with the present statistics.  
In fig. \ref{fig:d0k_prl} the fit to $m_{ES}$ distribution is shown for candidates respectively in the
double-Cabibbo suppressed mode (plot a), in the $D$ sidebands
(plot b) in the Cabibbo favored decay (plot c).
\begin{figure}[tb]
\begin{center}
\includegraphics[width=0.9\linewidth]{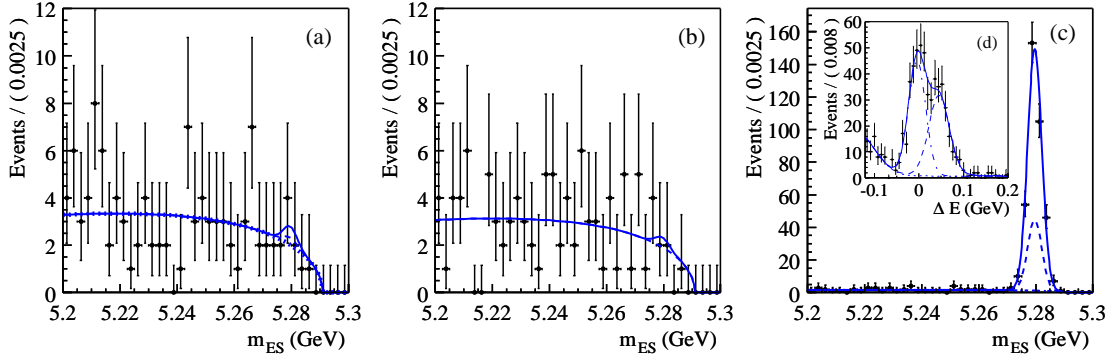}
\end{center}
\caption{$m_{es}$ distributions for (a) signal ($[K^\mp \pi^\pm]_D K^\pm$) 
candidates, (b) candidates from the $\dbarp$ sideband, and (c) $B \rightarrow DK$ candidates.
(d) $\Delta E$ distribution for $B \rightarrow DK$ candidates;  the peak
centered at $\approx$ 0.05 GeV is from $B \rightarrow D \pi$. The superimposed 
curves are described in the text.  In (c), the dashed Gaussian centered at
$m_B$ represents the $B \rightarrow D\pi$ contribution estimated from (d).} 
\label{fig:d0k_prl}
\end{figure}
The excess of events in the suppressed mode $N_{suppr}=1.1 \pm 3.0$ is compatible with zero 
while the number of candidates in the favored mode is $N_{fav}=261 \pm 22$. This result only 
allows to set a limit for $R_{DCS}: R_{DCS}<0.026$ at $90\%$ CL. This limit, making no 
assumptions about $\gamma$ or the overall strong phase difference $\delta$ translates into
a limit for the ratio of the $B$ decay amplitudes $r_B<0.22$ at $90\%$ CL.
This result is by itself relevant, because if $r_B$ is small, as this analysis suggests,
the suppression of the $b \rightarrow u$ amplitude will reduce the sensitivity  
of the ADS method in the measurement of $\gamma$ (see eq. \ref{eq:r_dcs}).   

\section{$B$ decays to states containing charmonium}
\subsection{Study of the process $B \rightarrow J/\psi K \pi \pi$ and search for $X(3872)$ }
One of the most recent results from \BABAR in the field of $B$ decays to states containing 
charmonium, is the study of the branching fraction of the process 
$B \rightarrow J/\psi K \pi\pi$. The analysis is even more important due to the observation 
of a state $X(3872) \rightarrow J/\psi \pi\pi$ by \mbox{Belle \cite{x3872_belle}} 
and \mbox{CDF \cite{x3872_cdf}}.
In addition, this analysis addresses the search for the unconfirmed $h_c(3526)$ charmonium state \cite{h_c}, 
and for an intrinsic charm component in the $B$ meson, leading to an anomalously large 
($B^- \rightarrow J/\psi D^0 \pi^-$) decay rate \cite{intrinsic}.
The study is based on the data collected by \BABAR during Run 1,2, and 3. As a first step, the 
energy substituted mass and $\Delta E$ distributions are reconstructed for $J/\psi K \pi \pi$
combinations (see fig. \ref{fig:jpsikpipi}). 

\begin{figure}[h]
\begin{center}
\begin{minipage}[t]{7cm}
\includegraphics[clip,width=7cm,height=7cm]{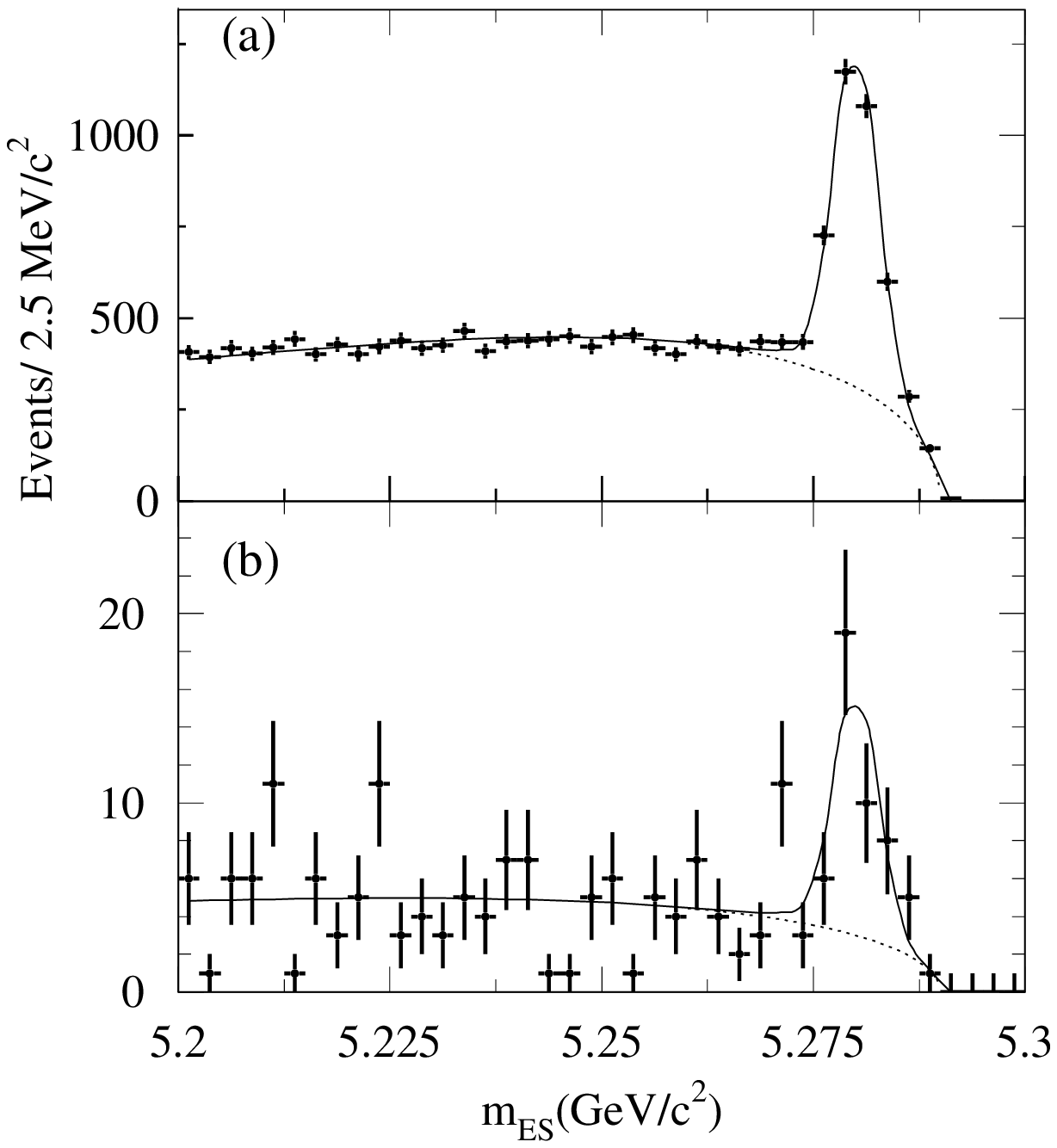}
\caption{Distribution of $m_{ES}$ for: \mbox{(a)$B\rightarrow J/\psi K \pi \pi$} candidates, 
(b) events
with \mbox{$3862 < m_{J/\psi\pi\pi} < 3882 MeV/c^2$}}
\label{fig:jpsikpipi}
\end{minipage}
\begin{minipage}[t]{0.5cm}{~}
\end{minipage}
\begin{minipage}[t]{7cm}
\includegraphics[clip,height=7cm]{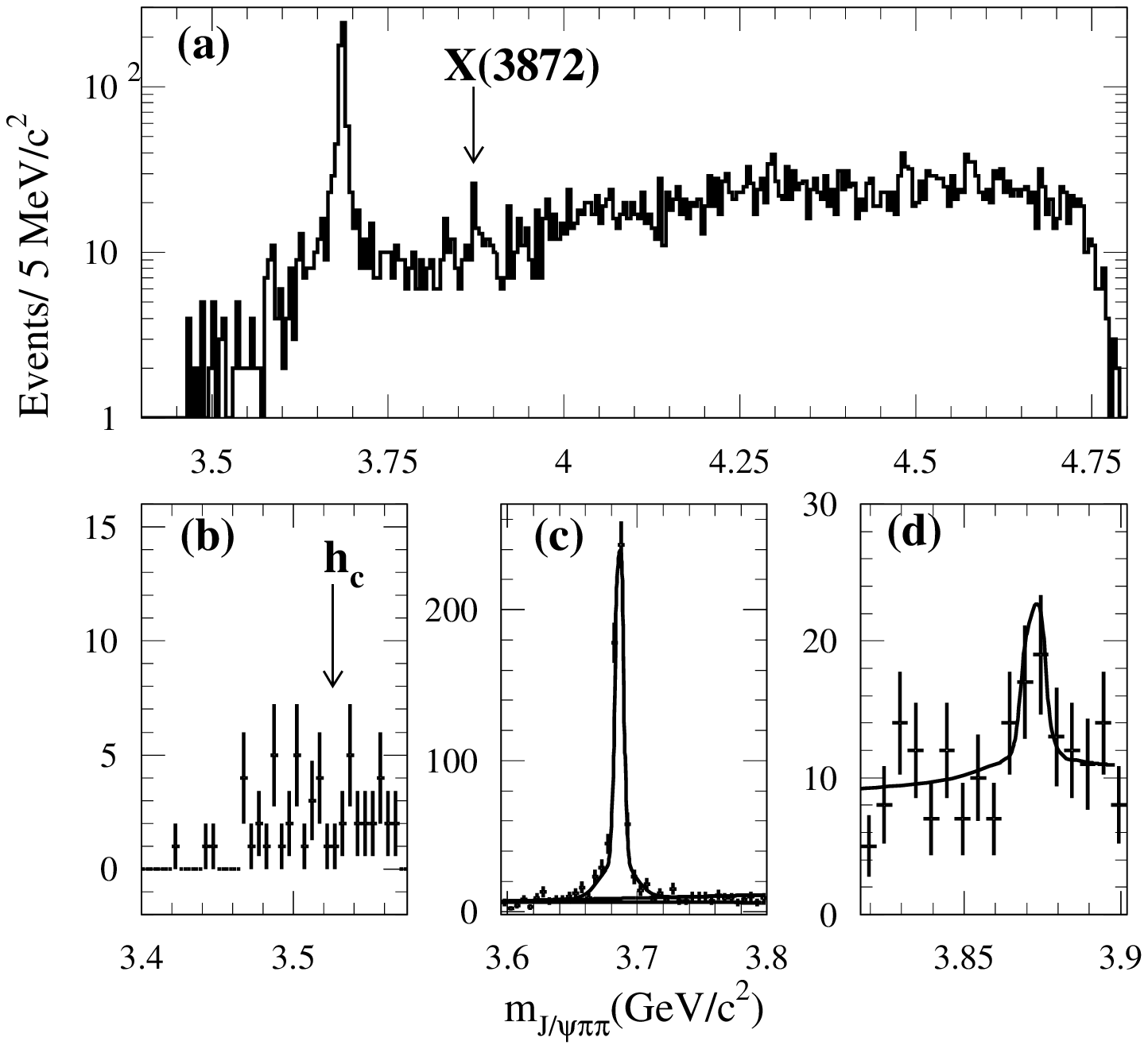}
\caption{Distribution of $m_{J/\psi\pi\pi}$ (top) and detailed analysis in the expected regions
for $h_c$ (bottom left), $\psi(2s)$ (bottom center), and $X(3872)$ (bottom right). No excess
is observed for the $h_c$ while a clear excee is visible in the region where the $X(3872)$ 
is expected.}
\label{fig:jpsipipi}
\end{minipage}
\end{center}
\end{figure}

\noindent  
There is a very clean signal in the expected region, with a yield of $N_{ev}=2540 \pm 72$ (upper plot), 
which allows to quote the branching fraction: 
$$BR(B\rightarrow J/\psi K \pi \pi)= (11.6 \pm 0.7 \pm 0.9) \times 10^{-4}$$  
If the same distribution is plotted for $J/\psi\pi\pi$ candidates inside a mass window of $20 MeV/c^2$
around the expected mass of the $X(3870)$ there is still an excess of events in the 
$m_{ES}$ distribution (fig. \ref {fig:jpsikpipi}, lower plot), The indication that part of the process
proceeds through the X(3870) state is clear from the study of the invariant mass distribution of 
the $J/\psi \pi\pi$ combinations (see fig. \ref{fig:jpsipipi}) while no hint of a $h_c$ signal is found.


\noindent 
The product branching fractions is found to be:
$$ BR(B^-\rightarrow X(3872)K^-) \times BR(X \rightarrow J/\psi\pi\pi)=(1.28\pm0.41)\times 10^{-5}$$
\noindent
while limits are set for the $h_c$ production and the $B^- \rightarrow J/\psi D^0 \pi^-$ decay rate:
$$ BR(B^-\rightarrow h_c K^-) \times BR(h_c \rightarrow J/\psi\pi\pi)< 4.3 \times 10^{-6};
~BR(B^-\rightarrow J/\psi D^0 \pi)< 5.2 \times 10^{-5} ~(90\% C.L.)$$

\subsection{Study of the process $B \rightarrow J/\psi \eta K$}
An analysis has recently addressed this decay mode, not only because
the branching fraction has never been measured by \BABAR before, but in particular  
to understand better the properties of the $X(3872)$ state. Indeed, if the state 
$X(3872)$ behaves like conventional charmonium, then we could expect a product 
branching fraction:
$$ BR(B^-\rightarrow X(3872)K^-) \times BR(X \rightarrow J/\psi\eta) \approx \mathcal{O}(3 \times 10^{-6})$$
For this reason both the decays $B^0 \rightarrow  J/\psi \eta K_s$ and $B^+ \rightarrow  J/\psi \eta K^+$ 
have been investigated. A clean signal has been measured in both processes and the following 
branching fractions have been quoted:
$$BR(B^+ \rightarrow  J/\psi \eta K^+)=(10.8\pm2.3\pm2.4)\times 10^{-5}; 
~B^0 \rightarrow  J/\psi \eta K_s=(8.4\pm2.6\pm2.7)\times 10^{-5}$$
The specific search for the decay to proceed through the $X(3872) \rightarrow J/\psi \eta$ transition
has been carried out. The $J/\psi \eta$ mass distribution has been studied for combinations inside
the $m_{ES}$, $\Delta E$ window for the $B$ decay. No excess of events is observed in the region
where a contribution coming from $X(3872)$ should be expected to appear. This allows to set a limit
for the product branching fraction:
$$ BR(B^-\rightarrow X(3872)K^-) \times BR(X \rightarrow J/\psi\eta) < 7.7 \times 10^{-6}~ (90\% C.L.)$$
which is anyway still compatible with the phenomenological expectations.

\section{Conclusions}
Thanks to the unprecedented number of $B$ pairs collected at the last-generation B-factories, 
in the last few years many new results have become available. In this paper we have presented 
only a few examples of the most recent analyses from the \BABAR Collaboration in the sector 
of $B$ transitions to final states containing charm. Of particular interest is the new observation
and branching fractions measurement for $B$ decays to several $D_{sJ}D^{(*)}$ states and the 
studies of the spin of $D_{sJ}(2460)$. The first results on the $B \rightarrow DK$ transition where 
the $D$ is searched in its double-Cabibbo suppressed decay modes, and the extraction of a new 
limit on $r_B$. The observation of the $X(3872)$ decay to $J/\psi \pi \pi$ and the study of $X(3872)$
properties.

\end{document}